\documentclass[aps,prl,preprint,groupedaddress]{revtex4}

\usepackage{graphicx}
\begin{document}

%\preprint{}

\title{The two-point resistance of a cobweb with a superconducting boundary }



\author{ Zhi-Zhong Tan%$^{{*}}$
\footnote{E-mail: tanz@ntu.edu.cn ; ~~ tanz@163.com}}
\address{ Department of Physics, Nantong University, Nantong 226007, China}

\author{ J. W. Essam
%$^{\dag\dag} $
\footnote{E-mail: j.essam@rhul.ac.uk}}
\address{ Department of Mathematics, Royal Holloway College, University of London, Egham, Surrey TW20 0EX, England.}

\author{ F. Y. Wu
%$^{}{\ddag}$
 \footnote{E-mail: fywu@neu.edu}}
\address{ Department of Physics, Northeastern University, Boston, MA 02115, USA}

\date{\today}

\begin{abstract}
We consider the problem of  two-point resistance on an $m\times n$ cobweb network with a superconducting boundary, which
is topologically equivalent to  a geographic globe.
  We deduce a concise formula for the resistance between any two nodes on the globe
using  a method of direct summation pioneered  by one of us [Z. Z. Tan, et al, J. Phys. A 46, 195202 (2013)].
This method contrasts  the Laplacian matrix approach which
is difficult   to apply to the geometry of a globe.
Our analysis gives the result directly  as a single summation.


\pagenumbering{arabic}

\end{abstract}


\pacs{ }

\maketitle


\section{1. Introduction}

A classic problem in electric circuit theory first studied by Kirchhoff \cite{kirch} more than 160 years ago is the computation of resistances in resistor networks.  Kirchhoff formulated the problem in terms of the Laplacian matrix of the network and
also  noted that the Laplacian also generates spanning trees. For the explicit
computation of two-point resistances, Venezian \cite{venezian} in 1994 considered the resistance between
 two arbitrary nodes using the method of superposition. In 2000 Cserti \cite{cserti}
evaluated the two-point resistance  using the lattice Green's function. Their
studies are confined to regular lattices of infinite size.

In 2004, one of us \cite{wu} formulated
a different approach and derived
an  expression for the two-point resistance in  arbitrary finite and infinite lattices
in terms of the eigenvalues and eigenvectors of the Laplacian matrix.
The Laplacian analysis has also been extended to impedance networks after a slight
modification of the formulation of  \cite{tzengwu}.
We shall refer to these methods as the {\it Laplacian} approach.
Applications of the Laplacian approach
 require a complete knowledge of the eigenvalues
and eigenvectors of the Laplacian  straightforward to obtain
for regular lattices.
 But it is generally difficult to solve the eigenvalue problem
  for non-regular networks
 such as a cobweb.

The cobweb
is a two-dimensional cylindrical network plus the insertion  of an additional
 node connected  to every node on one of the 2 boundaries.
An example of the cobweb is shown in the left panel of Fig. 1. In 2013 Tan, Zhou and Yang
\cite{tzy2013} proposed a conjecture, the TZY conjecture,  on the resistance between 2 nodes on
the cobweb.  It is then difficult to adopt the Laplacian approach directly to the problem
due to the special geometry of the cobweb.
However, by modifying the   method slightly to take care of the special cobweb geometry,
   Izmailian, Kennna and Wu (IKW) succeeded in establishing the TZY conjecture
using a modified Laplacian approach \cite{ikw}.

  In this paper we consider  the cobweb network with a superconducting boundary.  The
superconducting boundary of the cobweb
shrinks the boundary into one point resulting in a network of the shape of a ball, or a globe,
 shown in the right panel of Fig. 1.
An $m\times n$ cobweb network of $m$ rows and $n$ columns with a superconducting boundary is then
equivalent  to a globe with $m-1$ latitudes and $n$ longitudes. The example of $m=6, n=12$ is shown in Fig. 1.
 Since there  are 2  poles on a globe,  both the Laplacian  and
 the IKW modified Laplacian approaches are difficult to apply.

On the other hand, studies of the resistance problem had  been carried out independently by Tan and co-workers
 along a different route, which we shall refer to as the method of {\it direct} evaluation
\cite{tzy2013,tan,tanzhouluo13,tanchen13}.
 The direct method is  useful in cases when there exists a special node
such as a pole of the globe and the center of the cobweb,   connected to all other nodes
along lines such as the
  longitudes of a globe. This special connectivity
makes it possible to compute the resistance between  2 nodes by  computing separately their
relative potentials with respect  to the special node. One thus circumvents the need of diagonalizing
 a non-regular
Laplacian matrix.
The direct method of computing resistances was pioneered by one of us \cite{tan} and has been applied successively
 to the
cobweb network for  specific values of $m$ up to  $m=4$ \cite{tzy2013,tan,tanzhouluo13,tanchen13},
It has also been used recently to compute the resistances in a fan network \cite{essamtanwu}.
In this paper  we  apply the direct method  to the globe problem.

\begin{figure*}
\begin{center}
\includegraphics[width=10cm,bb=0 0 490 298]{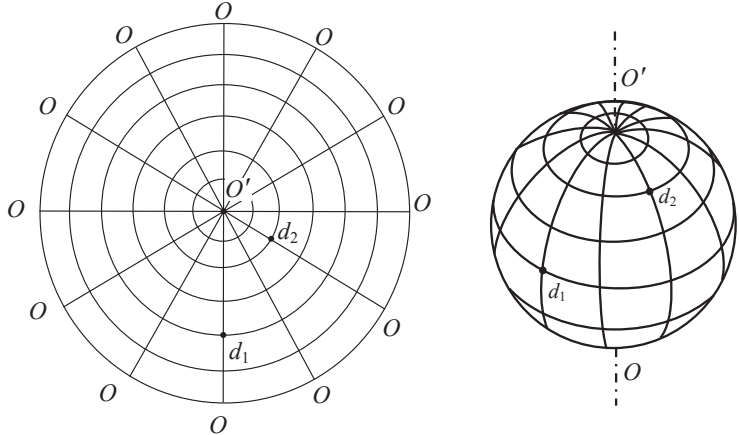}
\caption{ A $6\times 12$ cobweb network with a superconducting boundary and the equivalent globe with
5 latitudes and 12 longitudes.
  Bonds in longitude and latitude directions represent, respectively,  resistors $r_{0}$ and $r$.
The cobweb center is the north pole $O'$, and the boundary contracts into
south pole denoted by  $O$. }
\end{center}
\end{figure*}

\section{2. The equivalent resistance - the main result}

Consider the globe with $n$ longitudes and $m-1$ latitudes shown in Fig. 1.
Bonds in longitude and latitude directions have respective resistance $r_{0}$ and $r$ and
 let the south pole $O$ be the origin of coordinates.
Define variable $L_i$, and for later uses $ \lambda_i, {\bar \lambda}_i $ by
\begin{eqnarray}
 \lambda_i  \equiv e^{2L_i}&=& 1+h-h \cos \theta_i  + \sqrt{(1+h-h \cos \theta_i)^2-1}\nonumber \\
{\bar \lambda} _i  \equiv e^{-2L_i} &=& 1+h-h \cos \theta_i  - \sqrt{(1+h-h \cos \theta_i)^2-1} \nonumber\\
  \cosh 2L_i &=& 1 + h - h \cos \theta_i \label{Ldefinition}
\end{eqnarray}
where
\begin{eqnarray}
h=r/r_0, \quad \theta_{i}={(i-1)\pi}/{m}, \qquad  i=1,2, ... ,m . \nonumber
\end{eqnarray}
We find the resistance between the two nodes $d_1=\{1,y_1\} $ and $ d_{2}= \{x+1,y_2\}$, where $\{x,y\}$
 are coordinates, to be
 given by
 the expression
\begin{eqnarray}
&&R_{m\times n}^{globe}(\{1,y_1\},\{x+1,y_2\})=\frac{(y_{1}-y_{2})^{2}}{mn}r_{0} \nonumber \\
 &&\quad + \frac{r}{m}\sum_{i=2}^{m}\frac{ \cosh(nL_i) (\sin^{2}y_{1}\theta_{i}+\sin^{2}y_{2}\theta_{i})
-2   \cosh[(n-2x)L_i]  \sin (y_{1}\theta_{i})\,\sin (y_{2}\theta_{i})} {\sinh (2L_i)\sinh (nL_i)} .\label{mainresult}
\end{eqnarray}
Particularly, we have the special cases:

{\bf Case 1}. When $d_{1}$ and $d_{2}$ are on the same longitude at  $\{ 1,y_1\}$ and $\{1, y_2\}$, we have
 \begin{eqnarray}
R_{m\times n}^{long}(d_{1},d_{2})=\frac{(y_{1}-y_{2})^{2}}{mn}r_{0}+\frac{r}{m}\sum_{i=2}^{m}(\sin y_{1}\theta_{i}-\sin y_{2}\theta_{i})^{2}\bigg[ \frac{\coth(nL_i)}{ \sinh (2 L_i) } \bigg]. \label{case1}
\end{eqnarray}

{\bf Case 2}. When $d_{1}$ and $d_{2}$  are on the same latitude at $\{1,y\}$ and $\{x+1,y\}$, we have
\begin{eqnarray}
R_{m\times n}^{latt}(d_{1},d_{2})= \frac {4r}{m}\sum_{i=2}^{m}\frac { \sinh(xL_i) \sinh [(n-x)L_i] }
 { \sinh (2L_i) \sinh (nL_i) } \big[\sin^{2}(y\theta_{i})\big] , \label{case2}
\end{eqnarray}
The expression (\ref{case2}) is invariant under $ x \leftrightarrow  (n-x)$ as expected.

{\bf Case 3}. The resistance between a node at $\{x ,y\}$ and the north pole $O'$ is
\begin{equation}
R_{m\times n}(\{x, y\}, O') =
\frac{(m-y)^{2}}{mn}r_{0}+\frac{r}{m}\sum_{i=2}^{m}
\sin ^{2}(y\theta_i) \bigg[ \frac{\coth(nL_i)}{ \sinh (2 L_i) } \bigg]. \label{case3}
\end{equation}

{\bf Case 4}. The resistance between the two poles $O$ and $O'$ is
\begin{equation}
R_{m\times n} (O, O') = mr_0 / n.\label{case4}
\end{equation}

\section{3. Derivation  of the main result (\ref{mainresult})}
\subsection{3.1 Expressing the resistance in terms of longitudinal currents }
To compute the resistance between two nodes $ d_{1}=\{1,y_{1}\}$ and $ d_{2}=\{x+1,y_{2}\}$,
we inject a current $J$ into the network  at $d_{1}$ and exit the current  at $d_{2}$.
Denote the currents in all segments of the network as shown in Fig. 2. Then
by Ohm's law the potential differences  between $d_1$, $d_2$,
and the north pole $O'$ are, respectively,
\begin{eqnarray}
 U_{m\times n}^{globe}(d_{1},O')=r_{0}\sum_{i=y_{1}+1}^{m}I_{1}^{(i)},
 \qquad U_{m\times n}^{globe}(O', d_{2})=-r_{0}\sum_{i=y_{2}+1}^{m}I_{x+1}^{(i)},  \nonumber
\end{eqnarray}
where $I_1^{(i)}$ denotes currents along the longitude $1$, and $I_{x+1}^{(i)}$ denotes currents  along the
longitudinal $x+1$.
It then follows from the Ohm's law that  the resistance between $d_1$ and $d_2$ is
 \begin{eqnarray}
R_{m\times n}^{globe}(\{1,y_1\},\{x+1,y_2\})=\frac{r_{0}}{J}\Big[\sum_{i=y_{1}+1}^{m}I_{1}^{(i)}-\sum_{i=y_{2}+1}^{m}I_{x+1}^{(i)}\Big]. \label{R}
\end{eqnarray}
Therefore we need to find the longitudinal currents   $I_{1}^{(i)}$  and  $I_{x+1}^{(i)}$.
This is the main objective of this paper.

\begin{figure*}
\begin{center}
\includegraphics[width=9cm,bb=0 0 446 290]{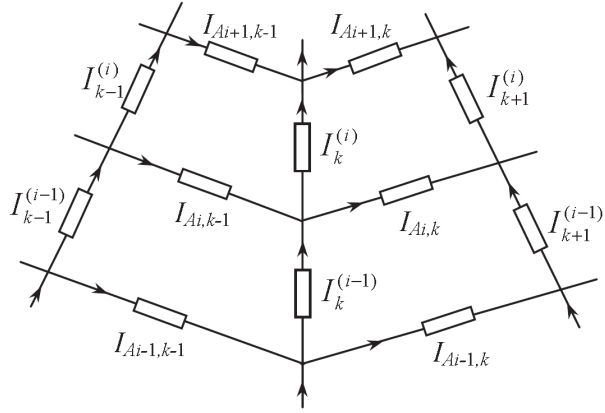}
\caption{ A segment of the globe  with  current directions.  }
\end{center}
\end{figure*}

\subsection{3.2 Matrix equation for longitudinal currents}
Analysis of the longitudinal currents is best carried out in terms of   a matrix equation.
Early discussions along this line
are due to Tan and co-workers \cite{tzy2013,tan,tanzhouluo13,tanchen13}.  A
similar analysis for a  fan network
 has been given recently  in  \cite{essamtanwu}.

A segment of the globe network is shown in Fig. 2
with  current labeling,  and we focus on the upper 2 rectangular meshes.
Around the 2 meshes  there are  5 longitudinal currents
$I^{(i)}_{k-1}, I^{(i)}_{k}, I^{(i)}_{k+1}, I^{(i-1)}_{k}, I^{(i+1)}_{k},$
and 4 horizontal currents $I_{Ai, k}$. The potential
across each current segment is  either $I_k^{(i)}r_0$ or $I_{Ai,k} r$.
The Kirchhoff law says that the sum of the potentials
around any closed loop is equal to zero. Apply this to the outer
perimeter of the two meshes, this gives a equation relating the 4 horizontal currents.
Furthermore,
the sum of all currents at a node must be zero.  Applying this Kirchhoff rule  to
the upper two consecutive nodes on the  longitude $k$, one
obtains 2 more equations relating the 4 horizontal currents.
However, it can be seen from Fig. 2 that the 4 horizontal currents enter all 3 equations only
in the combination of
 $ \Im _1 = I_{Ai+1, k-1} - I_{Ai+1, k} $ and  $\Im _2 = I_{Ai, k-1} - I_{Ai, k} $.
Thus one can eliminate $\Im _1$ and $\Im _2$ from the 3 equations.
 This gives the relation
\begin{equation}
I^{(i)}_{k+1} = - I^{(i)}_{k-1} +2(1+h) I^{(i)}_{k}  -h I^{(i+1)}_{k} - h I^{(i-1)}_k \label{longitudinal}
\end{equation}
  connecting the 5 longitudinal  currents.
After taking into account of modifications at $i=1,m$ \cite{essamtanwu},
  (\ref{longitudinal}) can be written
in a matrix form
\begin{eqnarray}
{\bf I}_{k+1}={\bf A}_{ m}{\bf I}_{k}-{\bf I}_{k-1}, \qquad  \label{matrixequation}
\end{eqnarray}
where   ${\bf A}_{m}$ and ${\bf I}_k$ are
 \begin{eqnarray}
{\bf A}_{ m}=
\left( {\begin{array}{cccccc}
   {{2+h}} & {{-h}} &{{0}}& 0& {\cdots}&{{0}} \\
   {{-h}} & {{2(1+h)}} & {{-h}} &0 &{\cdots} & {\cdots} \\
   {\vdots}&{\ddots} &{\ddots}&\ddots &{\ddots}&{\vdots}\\
   {{0}} & {\cdots}& 0 &{{-h}} & {{2(1+h)}} & {{-h}} \\
   {{0}} & {\cdots} &{{0}}& 0& {{-h}} & {{2+h}} \\
\end{array}} \right),
\qquad
{\bf I}_{ k}=
\left( {\begin{array}{cccccc}
     I_k^{(1)} \\ I_k^{(2)}  \\ {\vdots} \\ I_k^{(m-1)} \\ I_k^{(m)} \\
    \end{array}} \right).
\end{eqnarray}
It is understood that we have the  cyclic condition
\begin{equation}
{\bf I}_{0} = {\bf I}_n, \qquad  {\bf I}_{n+1} = {\bf I}_1. \label{cyclic}
\end{equation}
We consider the
 solution of (\ref{matrixequation}) in the next section.
\\

\subsection{3.3 General solution of the matrix equation}
   In this section we  consider
the solution of  (\ref{matrixequation}) in the absence of an injected current, namely, $J=0$.

The eigenvalues $t_i,\, i=1,2,...,m$ of ${\bf A}_{ m}$ are the $m$ solutions of
the  equation
\begin{equation}
{\rm det} \Big| {\bf A}_{ m} - t\  {\bf {\bar I}_{ m}} \Big| = 0, \label{eigenvalues}
\end{equation}
where ${\bf {\bar I}}_{ m}$ is the $m\times m$ identity matrix.
Since ${\bf A}_{ m}$ is Hermitian it can be diagonalized by a similarity transformation to yield
\begin{equation}
{\bf P}_{ m}\, {\bf A}_{ m}\, ({\bf P}_m)^{-1} = {\bf \Lambda}_{ m} \label{similarity}
\end{equation}
where ${\bf \Lambda}_{ m}$ is a diagonal matrix with eigenvalues  $t_i$  of ${\bf A}_m$ in the diagonal, and column vectors of  $({\bf P}_{ m})^{-1}$
are  eigenvectors of
${\bf A}_{ m}$.

It can be
verified that we have
 \begin{eqnarray}
 {\bf P}_{m}=
\left({\begin{array}{cccc}
   {1/\sqrt{2}} & {1/\sqrt{2}} &{\cdots} & {1/\sqrt{2}} \\
   {\cos(1-\frac{1}{2})\theta_{2}} & {\cos(2-\frac{1}{2})\theta_{2}} & {\cdots} & {\cos(m-\frac{1}{2})\theta_{2}} \\
   {\vdots} & {\vdots} & {\ddots}  &{\vdots} \\
   {\cos(1-\frac{1}{2})\theta_{m}} & {\cos(2-\frac{1}{2})\theta_{m}} & {\cdots} & {\cos(m-\frac{1}{2})\theta_{m}} \\
\end{array}} \right), \label{Pm}
\end{eqnarray}
\begin{eqnarray}
({\bf P}_{m})^{-1}=\frac{2}{m}
\left({\begin{array}{cccc}
   {1/\sqrt{2}} & {\cos(1-\frac{1}{2})\theta_{2}} &{\cdots} & {\cos(1-\frac{1}{2})\theta_{m}} \\
   {1/\sqrt{2}} & {\cos(2-\frac{1}{2})\theta_{2}} & {\cdots}  & {\cos(2-\frac{1}{2})\theta_{m}} \\
   {\vdots} & {\vdots} & {\ddots}   &{\vdots} \\
   {1/\sqrt{2}} & {\cos(m-\frac{1}{2})\theta_{2}} & {\cdots} & {\cos(m-\frac{1}{2})\theta_{m}} \\
\end{array}} \right) , \label{Pm-1}
\end{eqnarray}
where $ \theta_{i}={(i-1)\pi}/{m}$,
\begin{eqnarray}
t_{i}&=&2(1+h)-2h\cos\theta_{i}  = \lambda_i   + \bar {\lambda}_i  \nonumber \\
&=& 2 \cosh (2L_i), \quad\qquad\quad i=1,2,3,\cdots , m ,\label{ti}
\end{eqnarray}
where we have made use of  (\ref{Ldefinition}).

Apply ${\bf P}_{m}$ on the left  of (\ref{matrixequation}) and write
\begin{equation}
{\bf X}_{k} \equiv {\bf P}_{m}{\bf I}_{k},
\quad {\rm or} \quad {\bf I}_{k} = ({\bf P}_{m})^{-1} {\bf X}_{k} \label{IX} .
\end{equation}
After making use of (\ref{similarity}), we obtain  the equation
\begin{equation}
{\bf X}_{k+1}={\bf \Lambda}_{ m}{\bf X}_{k}-{\bf X}_{k-1} . \label{eigenvalue}
\end{equation}

Let the $i$-th element of the column vector ${\bf X}_{k}$ be $X_{k}^{(i)}$.
Then (\ref{eigenvalue}) gives
\begin{equation}
X_{k+1}^{(i)} = t_i X_{k}^{(i)} - X_{k-1}^{(i)} , \quad i = 1, 2, ..., m ,\label{recurrenteq}
\end{equation}
which is a set of recurrence relations for $ X_{k}^{(i)}$.

For $i=1$, the solution of (\ref{recurrenteq}),
which we shall make use later, is
particularly simply.  Since $\theta_1 = 0$ and $L_1 = 0$, we have  $t_1 =2$.
Then (\ref{recurrenteq}) becomes
\begin{eqnarray}
X_{k+1}^{(1)} = 2 X_{k}^{(1)} - X_{k-1}^{(1)} , \quad k = 1, 2, ..., n-1 ,\label{XXX}
\end{eqnarray}
which together with the cyclic condition
$X_0^{(1)} = X_n^{(1)}$ is a set of $n-1$ linear relations
for $n$ unknowns   $X_k^{(1)}, k = 1,2,...,n$, which is insufficient.
  But other than the trivial solution  $X_k^{(1)} =0$ which is useless,  we have also the obvious solution
that all  $X_k^{(1)}$'s are equal, namely,
\begin{equation}
 X_1^{(1)} = X_2^{(1)} = \cdots = X_n^{(1)}.  \label{XX}
\end{equation}

For $i>1$, the recurrence relation (\ref{recurrenteq})
can be solved by  the method of generating function.
Define  generating function
\begin{equation}
G(s) = \sum_{k=1}^\infty X_k^{(i)} s^k .\label{Gdefinition}
\end{equation}
 Multiply (\ref{recurrenteq}) by $s^k$ and sum both sides
of the equation from $k=1$ to $k=\infty$.  This yields
\begin{eqnarray}
\frac 1 s \Big[ G(s) - X_1^{(i)} s - X_2^{(i)} s^2 \Big]
   = t_i \Big[ G(s) - X_1^{(i)} s \Big] - s\,G(s)  \nonumber
\end{eqnarray}
from which we solve for $G(s)$, obtaining
\begin{equation}
G(s) = \frac {X_1^{(i)} s +\big(X_2^{(i)} - t_iX_1^{(i)} \big) s^2}
           {1-t_i s+ s^2}. \label{G}
\end{equation}

Partial fraction (\ref{G}) by using $1-t_i s+s^2 =(1- \lambda_i s)(1-{\bar\lambda}_i s)$ where
$\lambda_i $ and ${\bar \lambda}_i$ are defined in (\ref{Ldefinition}).  This gives
\begin{eqnarray}
\frac {1} {1-t_i s+ s^2 } = \frac 1 { \lambda_i - {\bar\lambda}_i}
     \bigg( \frac {\lambda_i }{ 1- \lambda_i s}  -  \frac {\bar\lambda_i }
                   { 1- {\bar \lambda}_i  s  }  \bigg),          \nonumber
\end{eqnarray}
which we substitute into (\ref{G}).
 Expand the right-hand side of  (\ref{G}) into a series in \ $s$ \   by making use of
  $(1-z)^{-1} = 1 + z +z^2 + \cdots$,
and compare both sides term by term.  We obtain after making use of the identity
$F_{k}^{(i)} - t_i F_{k-1}^{(i)} = - F_{k-2}^{(i)}$
 the solution of $ X_{k}^{(i)}$ in terms of a given initial condition of $X_{1}^{(i)}$ and $X_{2}^{(i)}$,
\begin{equation}
X_{k}^{(i)}=X_{2}^{(i)}F_{k-1}^{(i)} -X_{1}^{(i)}F_{k-2}^{(i)}, \quad\quad i > 1, \quad k \geq 1, \label{Xsolution}
\end{equation}
where
\begin{equation}
F_{k}^{(i)} =
\frac{\lambda_i^k - {\bar\lambda}_i^k}{\lambda_i - {\bar\lambda}_i}
   = \frac {\sinh (2k L_i)} {\sinh (2L_i)}.  \label{Fdefinition}
\end{equation}
In a similar fashion by  considering the generating function (\ref{Gdefinition}) with a summation
over k from $k=u +1$ to $\infty$ with a given initial condition of $X_{u+2}^{(i)}$ and $X_{u+1}^{(i)} $,
where $ u \geq 0$ is arbitrary, we obtain the solution
\begin{equation}
X_{k}^{(i)}=X_{u+2}^{(i)}F_{k-u-1}^{(i)} -X_{u+1}^{(i)}F_{k-u-2}^{(i)},
  \quad i>1, \quad u \geq 0,  \	\quad k \geq u+1 . \label{Xsolution1}
\end{equation}
Note that (\ref{Xsolution1})  reduces to (\ref{Xsolution}) when $u=0$.

\subsection{3.4 Boundary conditions with  input and output currents }
While either (\ref{Xsolution}) or (\ref{Xsolution1}) serves to determine ${\bf I}_k$ when there is no external
current injected to the network,
to compute the resistance between nodes  $d_1=d_1(1, y_1)$ and $d_2=d_2(x+1, y_2)$
we need to inject current $J$ at $d_1$ and exit the current at $d_2$. Then (\ref{Xsolution})
holds only for $1\leq k \leq x+1$.  For $k$ in the range of  $ x+1 \leq k \leq n+1$,
%(since $X_{n+1}^{(i)} = X_{1}^{(i)}$),
however, we need to use
(\ref{Xsolution1}) with $u=x$.
   Thus the injection of $J$ at $d_1(1, y_1)$ and the exit of $J$ at $d_2=d_2(x+1, y_2)$ specialize
(\ref{matrixequation}) for $k=1$ and $k= x+1$  to
\begin{eqnarray}
{\bf I}_2 &=& {\bf A}_m {\bf I}_1 - {\bf I}_n -J {\bf H}_1 ,\label{I2} \\
{\bf I}_{x+2}&=& {\bf A}_m {\bf I}_{x+1} - {\bf I}_x -J {\bf H}_2 ,\label{Ix}
\end{eqnarray}
where we have made use of the cyclic condition ${\bf I}_0 = {\bf I}_n$, ${\bf H}_1$ and ${\bf H}_{2}$ are
column matrices with elements
\begin{eqnarray}
(H_1)_i&=& h(-\delta_{i,y_1} + \delta_{i,y_1+1}) ,\nonumber \\
(H_2)_i &=& h(\delta_{i,y_2} - \delta_{i,y_2+1}), \nonumber
\end{eqnarray}
or, equivalently,
\begin{eqnarray}
{\bf H_{1}}&=&[\overbrace{0, \cdots0,-h,h}^{\mathrm{from} ~ 0\mathrm{th} ~ \mathrm{to}
~ (y_{1}+1)\mathrm{th}},0,\cdots,0\ ]^{T} , \nonumber \\
{\bf H_{2}}&=&[\overbrace{0, \cdots0,h,-h}^{\mathrm{from} ~  0\mathrm{th} ~ \mathrm{to}
~ (y_{2}+1)\mathrm{th}},0,\cdots,0]^{T} , \nonumber
\end{eqnarray}
where $[\ ]^T$
denote matrix transposes.

Applying ${\bf P}_m$ to (\ref{I2}) and (\ref{Ix}) on the left, we are led to
\begin{eqnarray}
{\bf X}_{2}&=& {\bf \Lambda}_m {\bf X}_{1}-{\bf X}_{n}-hJ{\bf D_{1}},  \label{D1}\\
{\bf X}_{x+2}&=& {\bf \Lambda}_m {\bf X}_{x+1}-{\bf X}_{x}-hJ{\bf D_{2}}, \label{D2}
\end{eqnarray}
where \ $h{\bf D}_1 = {\bf P}_m {\bf H}_1,\  h{\bf D}_2 = {\bf P}_m {\bf H}_2$, or equivalently, 
\begin{eqnarray}
  {\bf D_{1}}&=&[\zeta_{1,1},\zeta_{1,2},\cdots,\zeta_{1,i},\cdots,\zeta_{1,m-1},\zeta_{1,m}]^{T} \nonumber\\
  \zeta_{1,i}&=&P_{y_1,i} - P_{y_1+1,i}
    = -\cos\bigg(y_{1}-\frac{1}{2}\bigg)\theta_{i}+\cos\bigg(y_{1}+\frac{1}{2}\bigg)\theta_{i} \nonumber \\
   &=& -  2 \sin (y_1 \theta_i) \sin ( \theta_i /2),  \label{zeta1} \\
  {\bf D_{2}}&=&[\zeta_{2,1},\zeta_{2,2},\cdots,\zeta_{2,i},\cdots,\zeta_{2,m-1},\zeta_{2,m}]^{T}\nonumber\\
  \zeta_{2,i}&=&P_{y_2,i} - P_{y_{2}+1,i}
    = \cos\bigg(y_{2}-\frac{1}{2}\bigg)\theta_{i}-\cos\bigg(y_{2}+\frac{1}{2}\bigg)\theta_{i} \nonumber \\
   &=&   2 \sin \big(y_2 \theta_i) \sin ( \theta_i/2 ). \label{zeta2}
\end{eqnarray}
Explicitly,  (\ref{D1}) and (\ref{D2}) read
\begin{eqnarray}
X^{(i)}_2 = t_i X_1^{(i)} -X^{(i)}_n - hJ \zeta_{1,i}, \label{D11} \\
X^{(i)}_{x+2} = t_i X^{(i)}_{x+1} -X^{(i)}_x - hJ \zeta_{2,i},\label{D22}
\end{eqnarray}
where $t_i = 2 \cosh 2L_i $.

To determine the initial conditions  $X_{1}^{(i)}, X_{x+1}^{(i)} $ needed in our resistance
calculation (\ref{R}), we
 set $k=x, x+1$  in (\ref{Xsolution}),  $u=x$ and $k=n, n+1$  in (\ref{Xsolution1}) and making use of
the cyclic condition (\ref{cyclic}) $X_{n+1}^{(i)} = X_1^{(i)}$.   Together with
(\ref{D11}) and (\ref{D22}) this gives
 6 equations relating the 6 unknowns $X_1^{(i)}, X_2^{(i)}, X_n^{(i)}, X_x^{(i)}, X_{x+1}^{(i)}, X_{x+2}^{(i)}$,
\begin{eqnarray}
  \quad\left({\begin{array}{cccccc}
   F^{(i)}_{x-2} & -F^{(i)}_{x-1} & 0 & 1& 0 &0   \\
   F^{(i)}_{x-1} & -F^{(i)}_{x} & 0 & 0&1& 0\\
   t_i  & -1 &  -1& 0&0 &0  \\
    0& 0 &1& 0&  F^{(i)}_{n-x-2} & -F^{(i)}_{n-x-1}  \\
    1& 0 &0& 0&  F^{(i)}_{n-x-1} & -F^{(i)}_{n-x}  \\
     0& 0 &0& -1&  t_i & -1 \\
 \end{array}} \right)
 \left({\begin{array}{cccccc}
    X_1^{(i)}\\
    X_2^{(i)}\\
     X_n^{(i)} \\
       X_x^{(i)}\\
      X_{x+1}^{(i)}\\
      X_{x+2}^{(i)} \\
  \end{array}} \right)
     =
 \left({\begin{array}{cccccc}
    0\\
0\\
    hJ\zeta_{1,i} \\
  0\\
    0\\
   hJ\zeta_{2,i}
  \end{array}} \right), \quad i>1,  \label{equation}
\end{eqnarray}
where $t_i=2\cosh (2 L_i)$ and $F_k^{(i)} = \sinh (2kL_i) / \sinh (2L_i) $.

Solving (\ref{equation}), we obtain after some algebra and reduction the 2
solutions needed  in our resistance calculation (\ref{R}),
\begin{eqnarray}
X_{1}^{(i)}&=& \frac{(F_{n-x}^{(i)}+F_{x}^{(i)})\zeta_{2,i}+F_{n}^{(i)}\zeta_{1,i}}
 {4\sinh^2n L_i }(hJ)  \nonumber \\
  &=&  hJ\bigg[ \frac{(F_{n-x}^{(i)}+F_{x}^{(i)})\sin(y_2\theta_i)-F_{n}^{(i)}\sin(y_1\theta_i) }
 {2 \sinh^2 n L_i } \bigg]   \sin ( \theta_i/2 ) ,\quad i>1 , \label{X1i}  \\
 X_{x+1}^{(i)}&=& \frac{(F_{n-x}^{(i)}+F_{x}^{(i)})\zeta_{1,i}+F_{n}^{(i)}\zeta_{2,i}}
                 {4\sinh^2n L_i } (hJ) \nonumber \\
           &=&  hJ\bigg[ \frac{-(F_{n-x}^{(i)}+F_{x}^{(i)})\sin(y_1\theta_i)+F_{n}^{(i)}\sin(y_2\theta_i) }
 {2 \sinh^2n L_i } \bigg]   \sin ( \theta_i/2 )  ,  \quad i>1. \label{X2i}
\end{eqnarray}
For completeness, we also list the other 4 solutions of (\ref{equation}) although they are not needed in our calculation,
\begin{eqnarray}
  X_2^{(i)}  &=& {\frac{{(F_{x - 1}^{(i)}  + F_{n - x + 1}^{(i)} )\zeta _{2,i} + (F_{ 1}^{(i)} + F_{n - 1}^{(i)} )
\zeta _{1,i} }}{{4\sinh^2n L_i }}} (hJ), \nonumber\\
 X_n^{(i)}  &=& \frac{{(F_{x + 1}^{(i)}  + F_{n - x - 1}^{(i)} )\zeta _{2,i}  + (F_1^{(i)}
 + F_{n - 1}^{(i)} )\zeta _{1,i} }} {{4\sinh^2n L_i }}  (hJ), \nonumber \\
 X_x^{(i)}  &=& {\frac{{(F_{x - 1}^{(i)}  + F_{n - x + 1}^{(i)} )\zeta _{1,i}  + ( F_{1}^{(i)}+ F_{n - 1}^{(i)} )
\zeta _{2,i} }}
{{4\sinh^2n L_i }}} (hJ) , \nonumber \\
 X_{x + 2}^{(i)}  &=&  {\frac{{ (F_{1}^{(i)}+F_{n - 1}^{(i)} )\zeta _{2,i}  + (F_{x + 1}^{(i)}+F_{n - x - 1}^{(i)})\zeta _{1,i} }}{{4\sinh^2n L_i }}} ( hJ) . \nonumber
\end{eqnarray}

Solutions (\ref{X1i}) and (\ref{X2i}) are useful for $i>1$.
For $i=1$ (\ref{X1i}) and (\ref{X2i})  give the trivial solutions $X_{1}^{(1)}=X_{x+1}^{(1)}=0$.
But when $i=1$ we have $\zeta_{1,i}=\zeta_{2,i}=0$ so (\ref{D11}) and (\ref{D22})
reduce to (\ref{XXX}).  Then using the same argument leading to (\ref{XX}), we again obtain
$ X_1^{(1)} = X_2^{(1)} = \cdots = X_n^{(1)}$. This permits us
to write
\begin{eqnarray}
 X_1^{(1)}= \frac 1 n \sum_{k=1}^n  X_k^{(1)}=  \frac 1 n \sum_{k=1}^n \sum_{j=1}^m \big[ ({\bf P_m})_{1 j}I_{k}^{(j)}\big]
   = \frac 1 {\sqrt 2\,n}\sum _{i=1}^m \sum_{k=1}^n  I_k^{(i)} \label{totalsum}
\end{eqnarray}
where we have made use of   $({\bf P_m})_{1 j} = 1/\sqrt 2$.

The summations in (\ref{totalsum}) are taken over all longitudinal current segments on the globe.
Since the current $J$ flows from a node at latitude $y_1$ to a node at latitude $y_2$, by conservation
of current
 the summation over segments at a given latitude $i$ must yield $J$ for $y_1<i \leq y_2$ and zero otherwise, namely,
\begin{eqnarray}
\sum_{k=1}^n I_k^{(i)} &=& J, \qquad\quad y_1 < i < y_2+1 \nonumber \\
                       &=& 0, \qquad\quad {\rm otherwise} , \label{crosection}
\end{eqnarray}
so (\ref{totalsum}) gives the simple result
\begin{eqnarray}
X_1^{(1)} = \frac J {\sqrt 2\, n} (y_2-y_1). \label{x1solution}
\end{eqnarray}

\subsection{3.5 The equivalent resistance}
We are now in a position to  evaluate  the resistance  (\ref{R}).
From (\ref{IX}) we have
\begin{eqnarray}
I_1^{(i)} =  \sum_{j=1}^m [({\bf P}_m)^{-1}]_{ij} X_1 ^{(j)}. \nonumber
\end{eqnarray}
Using $({\bf P}_m)^{-1}$ given by (\ref{Pm-1}) with
$({\bf P}_m)^{-1})_{i1} = \sqrt 2 /m$ for all $i$, it is clear that the $j=1$ term in the summation needs to be singled out.
This gives
\begin{eqnarray}
I_1^{(i)} =  \ \frac  {\sqrt 2} { m} X_1^{(1) }+
\frac 2 m \sum_{j=2}^m   X_1 ^{(j)} \cos\bigg(i- \frac 1 2\bigg) \theta_j  \label{I1i}
\end{eqnarray}
and thus
\begin{eqnarray}
\sum _{i=y_1+1}^m I_1^{(i)} = \frac {\sqrt 2} m (m-y_1)X_1^{(1)}  - \frac 1 m \sum _{j=2}^m X_1 ^{(j)}
   \bigg[ \frac {\sin (y_1 \theta_j)} {\sin ( \frac 1 2 \theta _j) } \bigg], \label{I11}
\end{eqnarray}
where we have used the  formula
\begin{eqnarray}
\sum_{i =y+1}^m \cos \bigg(i - \frac 1 2 \bigg) \theta_j = - \bigg[\frac {\sin (y\theta_j)} {2 \sin (\frac 1 2 \theta_j ) }
\bigg]  \label{identity}
\end{eqnarray}
which can be established by using  the identity $\sum_{k=1}^n \cos (k-\frac 1 2)x = \sin (nx) / 2 \sin (x/2)$
%  holds for $\theta _j = (j-1)\pi /m$ and
\cite{gr}.

 Substituting (\ref{x1solution}) into (\ref{I11}), we obtain
\begin{eqnarray}
\sum _{i=y_1+1}^m I_1^{(i)} = \frac {J } {mn}  (m-y_1) (y_2-y_1) - \frac 1 m \sum _{j=2}^m X_1 ^{(j)}
   \bigg[ \frac {\sin (y_1 \theta_j)} {\sin ( \frac 1 2 \theta _j) } \bigg]. \label{1}
\end{eqnarray}
Similarly, we also obtain
\begin{eqnarray}
\sum _{i=y_2+1}^m I_{x+1}^{(i)} = \frac {J } {mn}  (m-y_2) (y_2-y_1) - \frac 1 m \sum _{j=2}^m X_{x+1} ^{(j)}
   \bigg[ \frac {\sin (y_2 \theta_j)} {\sin ( \frac 1 2 \theta _j) } \bigg]. \label{2}
\end{eqnarray}
Substituting (\ref{1}) and (\ref{2}) into (\ref{R}), we obtain
 \begin{eqnarray}
R_{m\times n}^{globe}(d_{1},d_{2})=\frac{r_{0}}{m}\bigg[\frac{(y_{2}-y_{1})^{2}}{n}+\frac{1}{J}\sum_{i=2}^{m}\frac{X_{x+1}^{(i)}\sin(y_{2}\theta_{i})-X_{1}^{(i)}\sin(y_{1}\theta_{i})}{\sin(\frac{1}{2}\theta_{i})}\bigg]. \label{Rtemp}
\end{eqnarray}
  Finally, we obtain our main result (\ref{mainresult}) by  further substituting
 $X_1 ^{(i)}$ and $X_{x+1} ^{(i)}$ from (\ref{X1i}) and (\ref{X2i}) into (\ref{Rtemp}).

\subsection{3.6 Special cases}
 Case 1: When $d_1=\{1, y_1 \}$  and $ d_2= \{1, y_2\}$ are on the same longitude, we take $x=0 $,
(\ref{mainresult}) reduces immediately to (\ref{case1}).

Case 2: When $d_1 = \{1, y\}$ and $d_2 = \{x+1,y\}$ are on the same latitude $y$,
(\ref{mainresult}) immediately
reduces to (\ref{case2}).

Case 3: The resistance between a node at $\{x, y\}$ and the north pole $O'$ is obtained by setting $y_1 = y,\, y_2 = m$
in (\ref{case1}). This gives (\ref{case3}).

Case 4: The resistance between the two poles is obtained by setting both $y_1 = 0,\, y_2 = m$ in (\ref{case1}).  This gives
$R_{m\times n} (O, O') = mr_0/n$.  This result can also be deduced by considering $R_{m\times n} (O, O')$
as connecting $n$ linear chains of resistance $mr_0$ each  in parallel, since by symmetry there are no currents in the horizontal
direction.

\subsection{3.7  A simple example }

As an example, we apply (\ref{mainresult}) to a  $2\times 4$ globe shown in Fig. 3.
  In this case the summation  in (\ref{mainresult}) has only one term $i=2$ with
$\theta_2 = \pi/2, m=2, \ n=4$, and
\begin{eqnarray}
 &&\cosh( 2L_2 ) = 1+h,  \quad  \cosh (4L_2) = 1+4h+2h^2 ,\nonumber \\
&& \sinh(2L_2)\sinh(4L_2) = 2h(1+h)(2+h) . \qquad  \nonumber
\end{eqnarray}
For the resistance between $O$ and $A$, we use (\ref{case3}) with $y=1$ and obtain
 \begin{eqnarray}
R_{2\times 4}(O,A)&=& \frac{1}{8}r_{0}+ \bigg(\frac{r}{2}\bigg)
{\frac{{\cosh (4L_i )}\sin^{2}\theta_{2}}{{\sinh (2L_i) \sinh (4L_i) }}}
= \frac {4+11h+5h^2}{8(1+h)(2+h)} r_0.  \nonumber
\end{eqnarray}
For the resistance between $A$ and $B$, we use (\ref{case2}) with $x=y=1$, and obtain
\begin{eqnarray}
 R_{2\times 4}(A,B)&=& \frac{{\cosh 4L_i - \cosh 2L_i }}{{\sinh (2L_i) \sinh (4L_i )}}(r\sin^{2}\theta_2)
=\frac{h(3+2h)}{2(1+h)(2+h)} r_0. \nonumber
\end{eqnarray}
For the resistance between $A$ and $C$, we use (\ref{case2}) with $x=2, y=1$, and obtain
\begin{eqnarray}
R_{2\times 4}(A,C) &=&{\frac{{\sinh ^2 (2L_i)}}  {{\sinh (2L_i) \sinh (4L_i) }}} (2r\sin^2 \theta_2 )
= \frac{h}{1+h}  r _0. \nonumber
\end{eqnarray}
The resistance between $O$ and $O'$ is given by (\ref{case4}) directly as
\begin{eqnarray}
R_{2\times 4}(O,O')= \frac{1}{2}r_{0}.   \nonumber
\end{eqnarray}
Here $A, B, C$ denote nodes shown in Fig. 3 and we have used  $r=hr_0$.  We have verified these results by carrying out explicit
 calculations.

\bigskip

\begin{figure*}
\begin{center}
\includegraphics[width=10cm, bb=0 0 416 271]{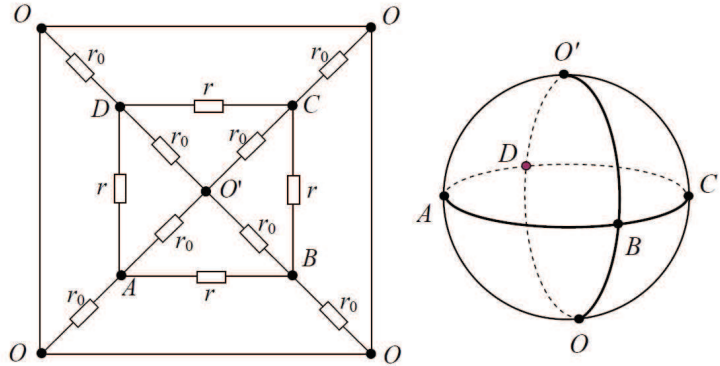}
\caption{ A $2\times 4$ globe and the associated cobweb network with a superconducting boundary.
 Node $O$ denotes the contraction of the superconducting boundary and is the coordinate center.}
\end{center}
\end{figure*}

\section{4.  Summary and discussion}
In 2004 Wu \cite{wu} established a theorem which  computes the equivalent resistance between
two nodes in a resistor network using the Laplacian approach. For the $m\times n$ network the results are  in the form of a double summation. Additional work is required to reduce this to a single summation.

An alternative direct approach of computing resistances had been developed
by Tan and co-workers \cite{tzy2013,tan,tanzhouluo13,tanchen13}
which, when applied to the cobweb and globe networks, gives the result  in terms of a single summation, thus offering
a direct and somewhat simpler approach. The direct method has been used by the present authors \cite{essamtanwu}
to deduce the 2-point resistance in a fan network. Here we use the direct method to compute resistances in a globe network, which is equivalent to the cobweb with a superconducting boundary. Our main result is (\ref{mainresult}) which gives
 the resistance between any two nodes of the globe.  Various special cases of the main result are also presented.

It is instructive to comment on why the Laplacian method is not used.
 While it  is tempting to use the Laplacian method and
formulate the globe problem as a cobweb with zero resistances along its boundary,
but
  since elements of the Laplacian are conductances, the inverse of resistances which is infinite, this is not easily done.
  It is  simpler and easier to use the direct approach.

Finally, we remark that the direct method of computing resistance can be extended to impedance networks, since
the Ohm's law based on which the method is formulated is applicable to impedances. This is advantageous than the
Laplacian method which needs to be modified when dealing with impedance networks as the Laplacian matrix is generally
 complex and non-Hermitian requiring special considerations \cite{tzengwu}.

\section*{Acknowledgment}
This work is supported by Jiangsu Province Education Science Plan Project (No. D/2013/01/048), the Research Project for Higher Education Research of Nantong University (No. 2012GJ003).


\section*{References}

\begin{thebibliography}{26}
\bibitem{kirch}
Kirchhoff G 1847 {\it Ann. Phys. Chem.} {\bf 72} 497
\bibitem{venezian}
G. Venezian, Am. J. Phys. {\bf 62} 1000 (1994).
\bibitem{cserti}  J. Cserti, Am. J. Phys. {\bf 68} 896 (2000).
%\bibitem{cserti}  Cserti J, D¨¢vid G and Attila Pir¨®th. Am. J. Phys.{\bf 70} ,153 (2002)
\bibitem{wu}  F. Y. Wu, J. Phys. A: Math. Gen. {\bf 37} 6653  (2004).
 \bibitem{tzengwu}  W. J. Tzeng,  F. Y. Wu, J. Phys. A: Math. Gen. {\bf 39} 8579 (2006).
% \bibitem{5}  N. Sh. Izmailian,M.C.Huang, Phys. Rev. E. 82 ,011125 (2010)
\bibitem{tzy2013}  Z. Z. Tan, L. Zhou and J. H. Yang, J. Phys. A: Math. Theor. {\bf 46} 195202 (2013)

\bibitem{ikw}  N. Sh. Izmailian, R. Kenna and F.Y.Wu. J. Phys. A: Math. Theor. {\bf 47} 035003(2014)
\bibitem{tan}  Z. Z. Tan, {\it Resistor network Models} (in Chinese) Xidian University of
   Science and Technology Press, Xian, China (2011).

\bibitem{tanzhouluo13}  Z. Z. Tan, L. Zhou and D. F. Luo, Int. J. Circ. Theor. Appl. DOI:10.1002/cta.1943 (2013).
\bibitem{tanchen13}  Z.Z. Tan, Q.H. Zhang ,  Int. J. Circ. Theor. Appl. DOI:10.1002/cta.1988 (2014).
\bibitem{essamtanwu}  J. W. Essam, Z. Z. Tan, F. Y. Wu, Resistance between two nodes in
general position on an $m\times n$ fan network, arXiv:1312.6727 [cond-mat.stat-mech].
\bibitem{gr} I. S. Gradshteyn and I. K. Ryzhik, {\it Table of Integrals, Series and Products}
(Academic Press, New York 1965),
1.342.4, p. 30.)
%
\end{thebibliography}
\end{document}